\documentclass[12pt]{JHEP3}

\pdfoutput=1
\usepackage{bm}
\usepackage{epsfig}
\usepackage{citesort}
\usepackage{graphicx}
\usepackage{multirow}

\def\be{\begin{equation}}
  \def\ee{\end{equation}}
\def\bea{\begin{eqnarray}}
  \def\eea{\end{eqnarray}}

\title{The Crossing Statistic: \\Dealing with Unknown Errors in the Dispersion of Type Ia Supernovae} 

\author{Arman Shafieloo\\
	Department of Physics, University of Oxford\\ 
	1 Keble Road, Oxford, OX1 3NP, UK\\
	\emph{and} \\
	Institute for the Early Universe, Ewha Womans University\\ 
	Seoul, 120-750, South Korea\\
	E-mail: \email{arman@ewha.ac.kr}}

\author{Timothy Clifton\\
	Astrophysics Department, University of Oxford\\
        Denys Wilkinson Building, Oxford, OX1 3RH, UK\\
	E-mail: \email{tclifton@astro.ox.ac.uk}}

\author{Pedro Ferreira\\
	Astrophysics Department, University of Oxford\\
        Denys Wilkinson Building, Oxford, OX1 3RH, UK\\
	E-mail: \email{p.ferreira1@physics.ox.ac.uk}}

     
\abstract{We propose a new statistic that has been designed to be used in
situations where the intrinsic dispersion of a data set is not
well known:  The Crossing Statistic.  This statistic is in general less
sensitive than $\chi^2$ to the intrinsic dispersion of the
data, and hence allows us to make progress in distinguishing between
different models using goodness of fit to the data even when the errors involved are poorly understood.
The proposed statistic makes use of the shape and trends of a
model's predictions in a quantifiable manner.  It is applicable to a
variety of circumstances, although we consider it to be
especially well suited to the task of distinguishing between
different cosmological models using type Ia supernovae.  We show
that this statistic can easily distinguish between different models in
cases where the $\chi^2$ statistic fails. We also show that the last
mode of the Crossing Statistic is identical to $\chi^2$, so that it can
be considered as a generalization of $\chi^2$. }

\begin{document}

\section{Introduction}                        
\label{sec:introduction}

The intrinsic dispersion of the data plays a crucial role in comparing
theoretical models to observations.   
If, for some reason, we do not know this dispersion, then evaluating
which  model best fits a given set of data points can be
particularly difficult.  This is the problem we face in cosmology when we attempt to make
inferences about cosmological models using 
type Ia Supernovae (SN Ia). 

SN Ia act to some degree like standardized candles, and are
widely used in cosmology to probe the expansion history of the
Universe, and hence to investigate the properties of dark
energy. Indeed, it is from observations of SN Ia  that the first direct evidence for an
accelerating universe was found~\cite{acceleration}, and although this
result has far reaching physical consequences, a complete understanding of the
physics of SN Ia is still lacking.  This lack of understanding is manifest in the
largely unaccounted for intrinsic dispersion of SN Ia, which affects
almost any subsequent statistical analysis that one attempts to perform~\cite{systematics}.  
Given that the intrinsic dispersion of SN Ia, $\sigma_{\rm int}$, typically constitutes a large
fraction of the total error on a data point, $\sigma_{i}$, this is a serious problem. 

One procedure that is often used to find the {\it a priori} unknown
intrinsic dispersion is to look for the value of
$\sigma_{\rm int}$ that gives a reduced $\chi^2$ of $1$ for a particular model,
and then to use this value to determine the likelihood of the data given
that model.  Such an approach does indeed
allow one to distinguish between different models using the likelihood
function, but at the expense of losing much of the original concept of
`goodness of fit' (which is the essence of a $\chi^2$ analysis). 
Rather than directly answering the question of which model
actually fits the data best, we are then left with answering the question of which model can be
made to give an ideal fit to the data by adding the smallest possible
error bars. This gives us no direct information about which model
best fits the data, as the error bars have been adjusted by
hand so that they all fit perfectly.  Furthermore, by treating error bars in this way
it becomes very difficult to detect any features that may be present in the data.

If we want to determine the goodness of fit of different models to the
data, we must therefore take a different approach.  Standard
statistics, such as $\chi^2$, however, are only reliable when the
assumed parameterization of the model is correct, and when the errors on the data are
properly estimated.  Given that the true nature of dark energy is
still not known, and that we have no reliable theoretical
derivation of $\sigma_{\rm int}$, the application of $\chi^2$ statistics
to the SN Ia data is not at all straightforward.  These problems persist
even when using non-parametric or model independent approaches
\cite{non-parametric}.  There have been extensive discussions in the
literature on using supernovae data for the purposes of model
selection~\cite{sn_anal}, and a number of problems have been
identified with using statistical methods in inappropriate
ways~\cite{eric_miquel08}.

To address these difficulties we propose a new statistic, that we
call {\it the Crossing Statistic}.  This statistic
is significantly less sensitive than $\chi^2$ to uncertainties in the
intrinsic dispersion, and can therefore be used more easily to check
the consistency between a 
given model and a data set with largely unknown errors.  The Crossing
Statistic does not compare two models directly, but rather determines the
probability of getting the observed data given a particular theoretical model. It works with the data
directly, and makes use of the shape and trends in a model's
predictions when comparing it with the data.

In the following we will discuss the concept of goodness of fit and
show how the $\chi^2$ statistic is sensitive to the size of unknown
errors, as well as how it fails to distinguish between different cosmological
models when errors are not prescribed in a definite manner. We will then
introduce the Crossing Statistic and show how it can be used
to distinguish between different cosmological models when the
standard $\chi^2$ analysis fails to do so. For simplicity, we will restrict
ourselves to four theoretical models: (i) a best fit flat $\Lambda$CDM
model, (ii) a smooth Lema\^{i}tre-Tolman-Bondi
void model with simultaneous big bang~\cite{CFL_void},
(iii) a flat $\Lambda$CDM model with $\Omega_{\rm 0m}=0.22$, and (iv)
an open, empty `Milne' universe.  We
will use the Constitution supernova data set~\cite{constitution}
that consists of 147 supernovae at low redshifts and 250 supernovae at
high redshifts.  This data set is a compilation of data from the
SuperNovae Legacy Survey~\cite{SNLS}, the ESSENCE
survey~\cite{ESSENCE}, and
the HST data set~\cite{HSTSN}, as well as some older data sets, and is
fitted for using the SALT light-curve fitter~\cite{SALT}.  We adjust the size of the error-bars in
this data set by considering additional intrinsic
errors (added quadratically). By comparing with $\chi^2$ we then show
that the Crossing Statistic is relatively insensitive to the unknown
intrinsic error, as well as being more reliable in distinguishing
between different cosmological models. In a companion paper, we will
test a number of other dark energy models using this statistic.

\section{Method and Analysis}                        
\label{meth}
First let us consider the $\chi^2$ statistic. For a given data set
$(\mu^{\rm e}_{i}, i=1\cdots N)$ we have that $\chi^2$ is given by
\begin{eqnarray}
\chi^2=\sum_{i}^{\rm N}\frac{(\mu^{\rm t}_{i}-\mu^{\rm e}_{i})^2}{\sigma_{i}^2},
\end{eqnarray}
where $\mu^{\rm t}_{i}$ is the prediction of the model that we are comparing the data set to, and $\sigma_{i}$ are the
corresponding variances ($\sigma$ has units of magnitudes throughout).  If the data-points are uncorrelated and have
a Gaussian distribution around the distribution mean,
then we have a $\chi^2$ distribution with $N-N_{\rm P}$ degrees of freedom (where $N_{\rm P}$ is the number of parameters
in the theoretical model). 

\begin{figure}
\hspace{30 mm}
\includegraphics[width=100mm,height=75mm,angle=0]{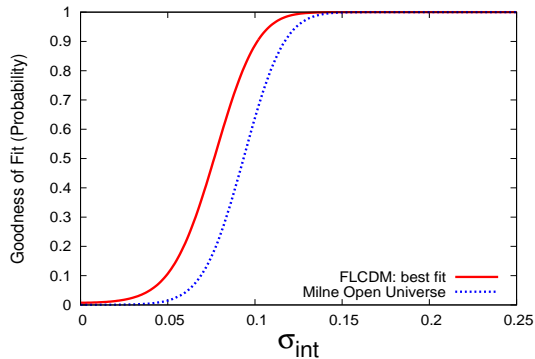}
\hfil
\caption{\small The $\chi^2$ goodness of fit of the Constitution supernova
  data~\cite{constitution} to a flat $\Lambda$CDM model (red
  line) and a Milne universe (blue line), assuming additional  
  intrinsic errors added quadratically to the errors specified in the
  data set. The $\chi^2$ goodness of fit for these two models can be seen to be
  comparable for different values of additional intrinsic error,
  making them difficult to distinguish without any knowledge of the
  value of $\sigma_{\rm int}$.}   
\label{fig:gof}
\end{figure}

Now let us now calculate the $\chi^2$ goodness of fit for two of our cosmological models:
a flat best fit $\Lambda$CDM model, and a Milne universe.  Let us also
assume an additional intrinsic error, $\sigma_{\rm int}$, on top of the error
prescribed in the Constitution data set, $\sigma_{i {\rm (data)}}$, so that
the total error is $\sigma^2_i=\sigma_{i {\rm (data)}}^2+\sigma_{\rm int}^2$. This will allow us to check
how sensitive our analysis is to coherent changes in the size of error
bars. In
Fig.~\ref{fig:gof} we plot the $\chi^2$ goodness of
fit for our two theoretical models as a function of
$\sigma_{\rm int}$. It can be seen that these two models cannot be
easily distinguished from each other using $\chi^2$
alone, unless the additional intrinsic error is already known.
We also note that the $\chi^2$ goodness of fit for the standard
flat $\Lambda$CDM model, given the Constitution data without any
additional intrinsic errors, is less than $0.6 \%$
($\chi^2=465.5$ for $397$ data points).  

\begin{figure}
\includegraphics[width=95mm,height=70mm,angle=0]{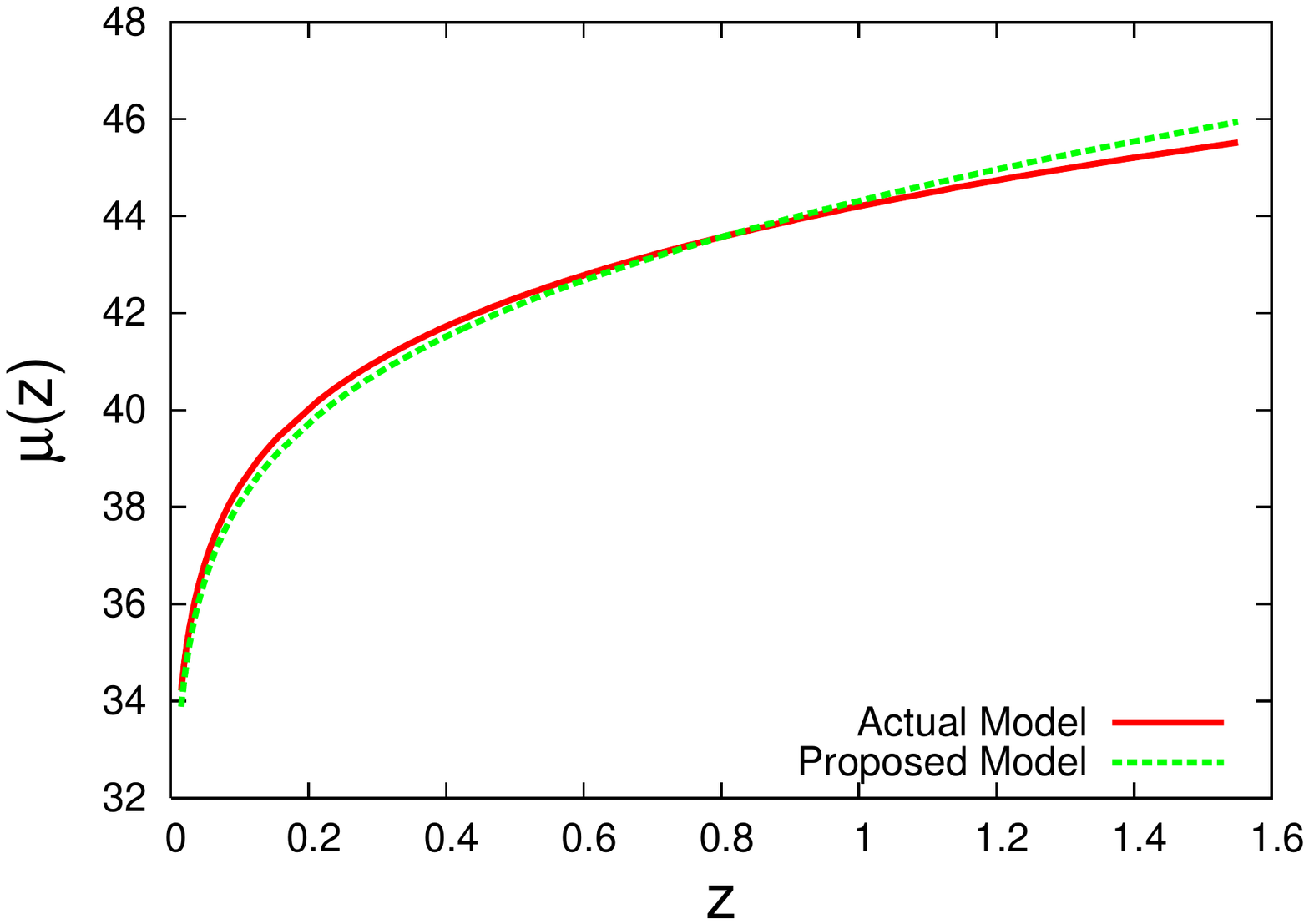}
\hfil
\hspace{-20mm}
\includegraphics[width=95mm,height=70mm,angle=0]{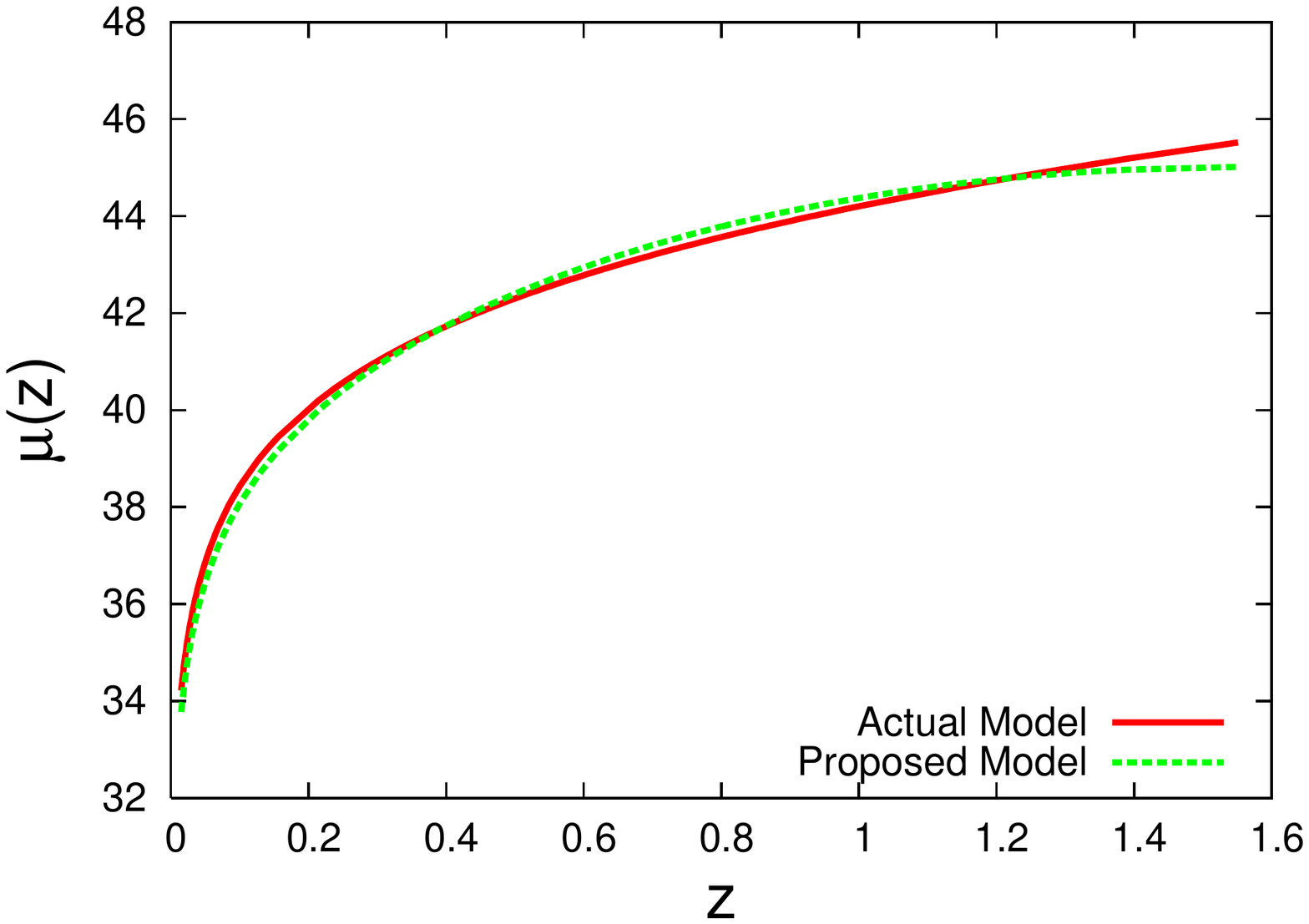}
\hfil
\caption{\small An idealized schematic plot of one crossing (left panel) and two crossings
  (right panel) between a proposed theoretical model and the actual
  model of the Universe when comparing magnitudes as a function of redshift, $\mu (z)$. In reality the
  actual Universe is observed in the form of data with error bars, of course.} 
\label{fig:schematic}
\end{figure}

If the real Universe differs from the assumed theoretical model, one
would hope that it would be possible to develop a statistical test
that would be able to pick up on this.  To these ends we consider the
`crossings' between the predictions of a given model, and the real
Universe from which the data has been derived. Figure \ref{fig:schematic} shows a
schematic picture of what we mean by one crossing (left panel), and two crossings (right
panel). In what follows we will use the existence of this type of crossing to develop
a new statistic that can be used to determine the goodness of fit between an assumed model and
the real Universe.


To build our Crossing Statistic in the case of SN Ia data, we must
 first pick a theoretical or phenomenological model of dark energy (e.g.
 $\Lambda$CDM) and a data set of SN Ia distance moduli $\mu_{i}(z_{i})$
 (e.g. the Constitution data set~\cite{constitution}). As in \cite{BND},
 we use the $\chi^2$ statistic to find the best fit form of the
 assumed model, and from this we then construct the {\it error normalized difference} of the data
 from the best fit distance modulus $\bar \mu(z)$:
\begin{equation} 
q_i(z_i)=\frac{\mu_i(z_i) - {\bar \mu(z_i)}}{\sigma_i (z_i)}. \label{qidef}
\end{equation}
Let us now consider the {\it one-point Crossing Statistic}, which  
tests for a model and a data set that cross each other at only one
point. We must first try to find this crossing point, which we label by $n_1^{\rm CI}$ and
$z_1^{\rm CI}$.  To achieve this we define
\begin{equation}
 T(n_1)=Q_1(n_1)^2+Q_2(n_1)^2,
\label{qndef} 
\end{equation}   
where $Q_1(n_1)$ and  $Q_2(n_1)$ are given by
\begin{eqnarray}
 Q_1(n_1)&=&\sum_{i=1}^{\rm n_1} q_i(z_i)    \nonumber \\
 Q_2(n_1)&=&\sum_{i=n_1+1}^{\rm N} q_i(z_i),
\end{eqnarray}
and where $N$ is the total number of data points.
If $n_1$ is allowed to take any value from $1$ to $N$ (when the data is sorted by redshift)
then we can maximize $T(n_1)$ by varying with respect to $n_1^{\rm CI}$.
We then write the maximized value of $T(n_1)$ as $T_{\rm I}$.
Finally, we can use Monte Carlo simulations to find how often we
should expect to obtain a $T_{\rm I}^{\rm MC}$ larger or equal to the value derived from the
observed data, $T_{\rm I}^{\rm data}$.
This information can then be used to estimate the probability that the
particular data set we have in our possession should be realized from the
cosmological model we have been considering.

In our analysis, the process of estimating the distribution of
$T_{\rm I}^{\rm MC}$ using Monte Carlo simulations is done in a model
independent way as follows. Firstly, a number of different data sets
are generated from a single fiducial model, which we take to
represent the `true' model of the Universe.  The residuals of the fake
data are then calculated by subtracting the mean values of the same
fiducial model, from which we can then determine $T_{\rm I}^{\rm MC}$.  As such, it follows
that $T_{\rm I}^{\rm MC}$ does not depend on the background model (which
is subtracted away from the generated data to find the residual), but rather
on the dispersion about the fiducial model that we haven chosen to
adopt.  This dispersion is taken to correspond to the errors on the
observational data, and so is itself model independent (up to the
extent that observers make assumptions about the background cosmology
when specifying their value).

Now, before applying the Crossing Statistic to real data, let us first
we consider how it fares when applied to simulations.  For this we create 1000 realizations of
data similar to the Constitution supernova sample based on a fiducial
flat $\Lambda$CDM model with $\Omega_{\rm 0m}=0.27$.  We then test
two different models using the same fake data sets.  The first of these is the fiducial
model itself (the `correct model'), and the second is a flat $\Lambda$CDM model
with $\Omega_{\rm 0m}=0.22$ (the `incorrect model').  These two models are
intentionally chosen to be similar to each other in order to
explicitly show the effectiveness of the Crossing Statistic at
distinguishing between different models.  Next, we add an extra intrinsic
dispersion of $\sigma_{\rm int}=0.05$ to the data and test the
two models again.  This is done to simulate the more realistic
situation in which the precise value of $\sigma$ is unknown.  Using the simulated data, and
applying our statistic, we then test how often the simulated data
is sufficient to rule out each of the two models at the $99\%$
confidence level (CL).  This data is displayed in Table~\ref{table:data},
along with the result of using $\chi^2$ alone\footnote{We call
  our statistic $T_{\rm I} + \chi^2$ in Table~\ref{table:data}, as we
  minimize for $\chi^2$ first, by adjusting the nuisance parameters,
  before calculating $T_{\rm I}$.}. 

\begin{table*}
\begin{tabular}{ccccccccc}
\\
 $$ & $\sigma_{\rm int}=0.0$ & $$ & $\sigma_{\rm int}=0.05$ & $$ & $$\\
\hline
 $$ & $T_{\rm I}$+$\chi^2$ & $\chi^2$ & $T_{\rm I}$+$\chi^2$ & $\chi^2$\\
\hline
Correct Model ($\Lambda$CDM with $\Omega_{\rm 0m}=0.27$) & $1\%$ & $1\%$ & $0.5\%$ & $0\%$ \\
\hline
Incorrect Model ($\Lambda$CDM with $\Omega_{\rm 0m}=0.22$) & $28.5\%$ & $1.9\%$ & $26.4\%$ & $0\%$\\
\hline
\\
\end{tabular}
\caption{A comparison of the $\chi^2$ and $T_{\rm I}$ statistics
  using data simulated from a $\Lambda$CDM model with
  $\Omega_{\rm 0m}=0.27$.  Percentages show the fraction of simulations in
  which the model in question is ruled out at the $99\%$ confidence level.}
\label{table:data}
\end{table*}

It can be seen that with $\sigma_{\rm int}=0$ the $\chi^2$ and $T_{\rm
  I} + \chi^2$ statistics both rule out the correct model at $99\%$ CL in
$1\%$ of the cases, as should be expected to happen from their definitions. The
  incorrect model, however, is ruled out by the $\chi^2$ statistic
at $99\%$ CL in less than $2\%$ of the cases only, while the $T_{\rm
  I} + \chi^2$ statistic is ruled out at $99\%$ CL
about $28.5\%$ of the time. This is a significant improvement in
distinguishing different models by using a more sophisticated
statistic that is extracting more information from the data. 
Also, when $\sigma_{\rm int}=0.05$ we can see that $\chi^2 + T_{\rm I}$
is still sensitive to the incorrect model, picking it up and ruling it out
at $99\%$ CL in about $26.4\%$ of cases.  This is not true of
$\chi^2$, and clearly demonstrates that $T_{\rm I}$ is much less sensitive to
the unknown value of $\sigma_{\rm int}$ than $\chi^2$, while being better at
distinguishing the correct model from the incorrect one.  In fact,
even if we over-estimate the size of the error-bars, $T_{\rm I}$ still
performs well, and frequently picks out the incorrect model with high
confidence.


\begin{figure}
\vspace{-10pt}
\includegraphics[width=95mm,height=70mm, angle=0]{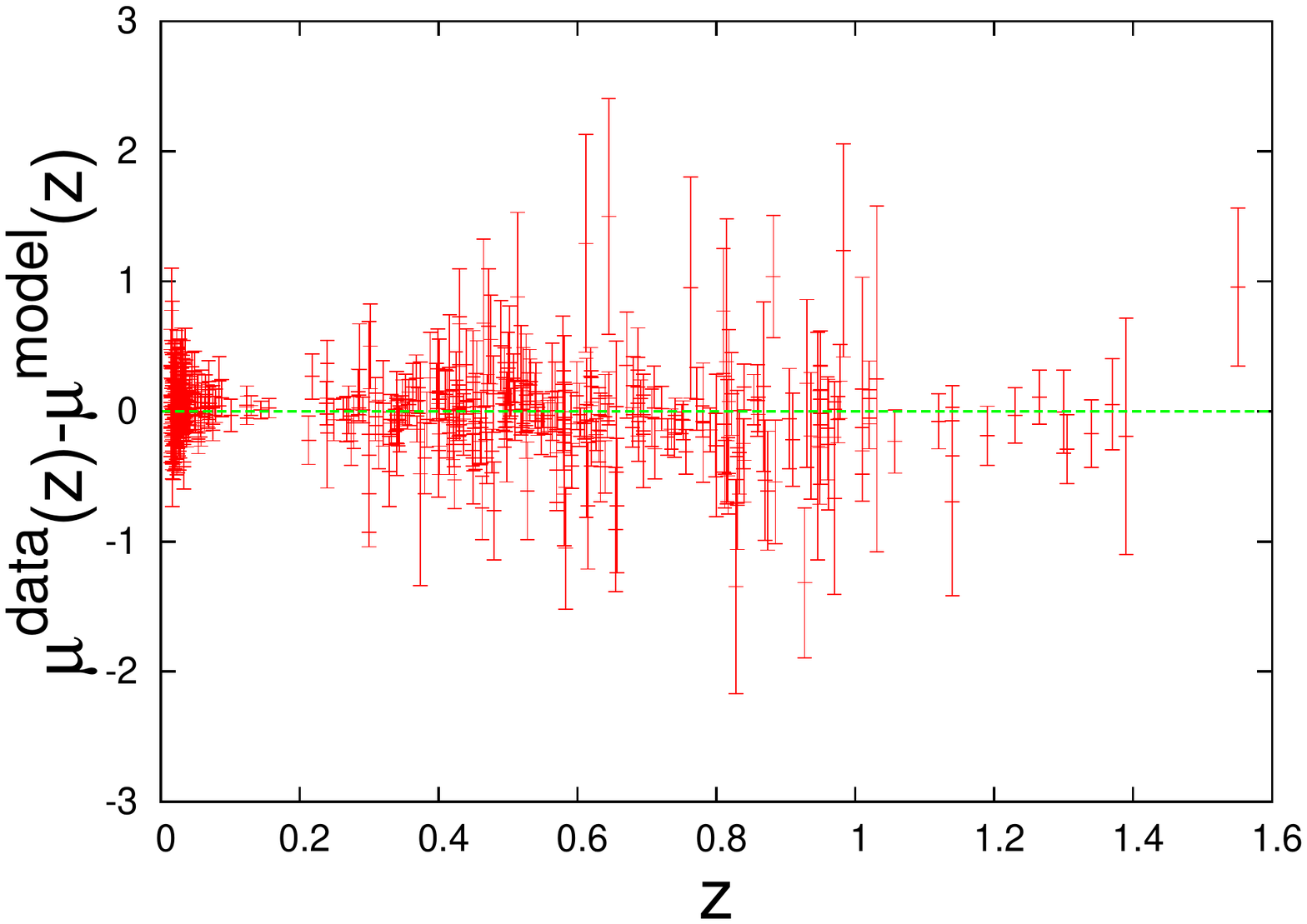}
\hspace{-20mm}
\includegraphics[width=95mm,height=70mm, angle=0]{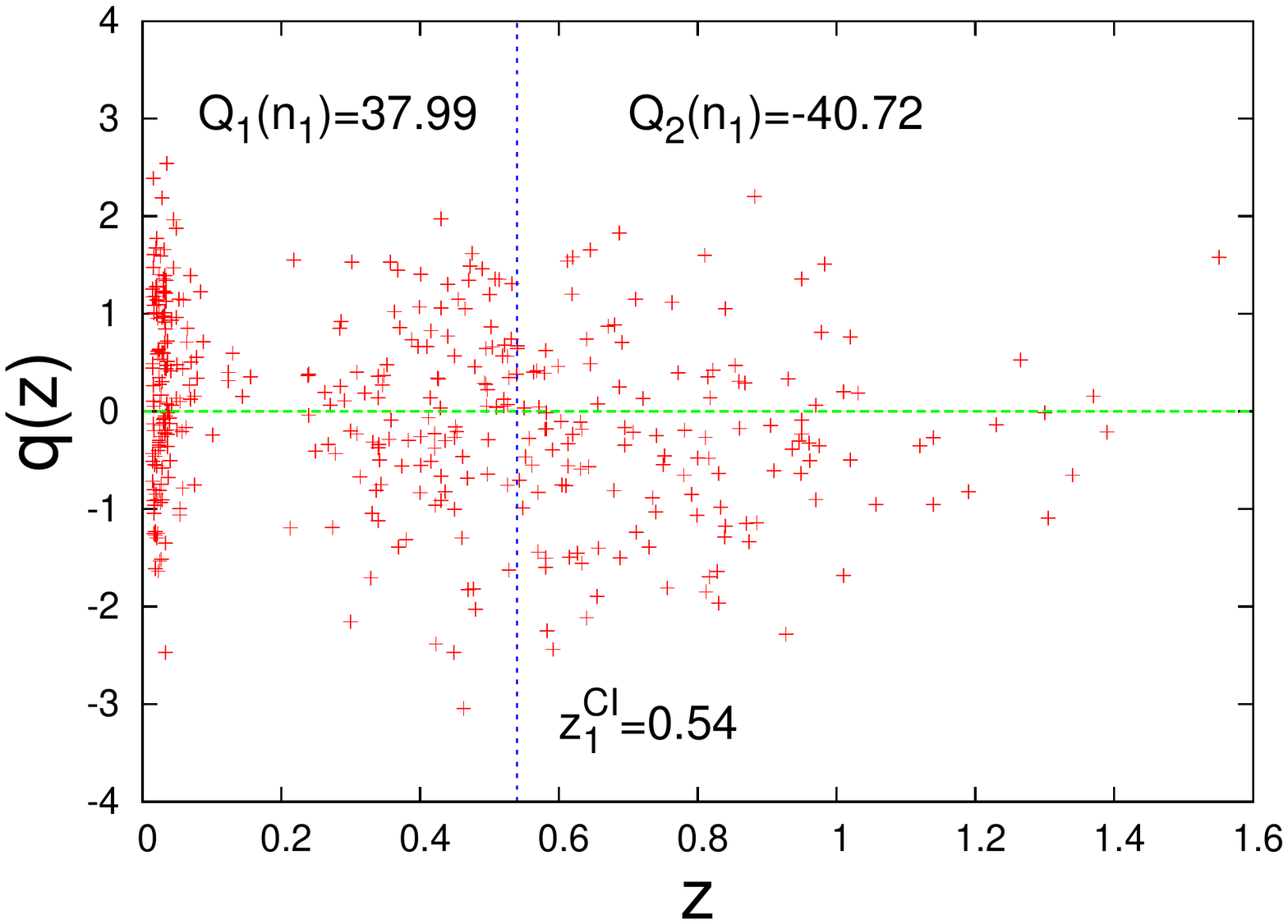}
\caption{\small Residuals with error bars taken from the data (left
  panel) and the error normalized difference, $q(z)$, (right panel),
  for a single random realization of supernova data similar to
  Constitution sample. The simulated data here is based on the
  fiducial model of flat $\Lambda$CDM with $\Omega_{\rm 0m}=0.27$,
  and the test model is flat $\Lambda$CDM
  with $\Omega_{\rm 0m}=0.22$. We assumed here that there is no extra intrinsic dispersion. The crossing point occurs at $z_1^{\rm CI}$, and is shown by the vertical blue dashed line.  Derived values of
  $Q_1(n_1^{\rm CI})$ and  $Q_2(n_1^{\rm CI})$ are also
  displayed. In the right-hand panel one can see the 
  unbalanced distribution of points around the green zero line (on
  the right side there are more points below the line, while on
  the left there are more points above it).} 
\label{fig:new}
\end{figure}

To elaborate further on why $\chi^2$ is often not sensitive to using
the incorrect model, while $\chi^2 +T_{\rm I}$ is, let us consider the
distribution of residuals with redshift.  This is shown in
Fig.~\ref{fig:new} for a single random realization of data generated
from a flat $\Lambda$CDM model with $\Omega_{\rm 0m}=0.27$, and using a
test $\Lambda$CDM model that is also flat with
$\Omega_{\rm 0m}=0.22$.  The distribution of the fake data points is
similar to that of the Constitution sample and the data has no extra intrinsic dispersion.
The green horizontal dashed line in Fig.~\ref{fig:new} is the
zero line about which the normalized residuals should fluctuate, 
when the model being tested and the actual model are the same. The
blue dotted vertical line in the right-hand plot represents the redshift at
which $T(n_1)$ is maximized, $z_1^{\rm CI}$.  The derived values of
$Q_1(n_1^{\rm CI})$ and  $Q_2(n_1^{\rm CI})$ on either side of
the blue line are also displayed.  For this particular realization of
the data the derived $\chi^2$ for the test model with
$\Omega_{\rm 0m}=0.22$ is $375.72$, which represents a very good $\chi^2$
fit to the data considering the number of data points is $557$.  The
corresponding P-value\footnote{P-value is defined as the probability that, given the null hypothesis, the value
of the statistic is larger than the one observed. We remark that in
defining this statistic one has to be cautious about {\it a
  posteriori} interpretations of the data.  That is, a particular feature
observed in the real data may be very unlikely (and lead to a low
P-value), but the probability of observing {\em some} feature may be
quite large -- see the discussion in \cite{Hamann:2009bz}.} derived from Monte Carlo simulations is more
than $50\%$. However, the
  derived value of $T_{\rm I}$ is $3102.13$, which comparing with the Monte
  Carlo realizations results in a P-value of less than $0.5\%$. This
  shows that the model with $\Omega_{\rm 0m}=0.22$ is strongly ruled out
  with the $\chi^2+T_{\rm I}$ statistic, at the level of $3\sigma$.

\begin{figure}
\hspace{30 mm}
\includegraphics[width=100mm,height=75mm,angle=0]{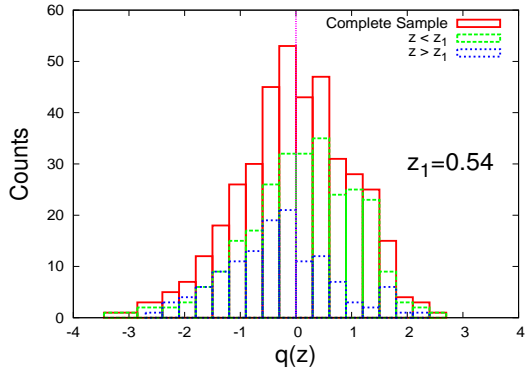}
\hfil
\caption{\small The distribution of error normalized difference
at different redshifts, as studied in Fig 3. While
the overall distribution has a reasonable Gaussian shape, the
normalized residuals at $z<0.54$ have a clear shift to the right
while those at $z>0.54$ have a clear shift to the left.}   
\label{fig:histo}
\end{figure}

This approach can be extended to models with more than one crossing
point by the {\it two-point Crossing Statistic}. In this case we
assume that the model and the data cross each other at two points and,
as above, we try to find the two crossing points and their red shifts,
which we now label $n_1^{\rm CII},z_1^{\rm CII}$ and $n_2^{\rm CII},z_2^{\rm CII}$.
This is achieved by defining
\begin{equation}
 T(n_1,n_2)=Q_1(n_1,n_2)^2+Q_2(n_1,n_2)^2+Q_3(n_1,n_2)^2,
 \end{equation}   
where the $Q_i(n_1,n_2)$ are now given by
\begin{eqnarray}
 Q_1(n_1,n_2) &=&\sum_{i=1}^{\rm n_1} q_i(z_i) \nonumber \\
 Q_2(n_1,n_2) &=&\sum_{i=n_1+1}^{\rm n_2} q_i(z_i) \nonumber \\
 Q_3(n_1,n_2) &=&\sum_{i=n_2+1}^{\rm N} q_i(z_i).
 \label{eq:QT2} 
\end{eqnarray}
We can then maximise $T(n_1,n_2)$ by varying with respect to $n_1$
and $n_2$, to get $T_{\rm II}$.  Comparing $T_{\rm II}$ with the results from
Monte Carlo realizations then allows us to determine how often
we should expect a two-point crossing statistic that is greater than
or equal to the $T_{\rm II}$ obtained from real data.  The {\it three-point
  Crossing Statistic}, and higher statistics, can be defined in a
similar manner.  This can continue up to the {\it N-point Crossing
  Statistic} which is, in fact, identical to $\chi^2$.  We
also note that the zero-point Crossing Statistic, $T_{0}=(\sum_{i}^{\rm N} q_i)^2$, 
is very similar to the Median Statistic developed by Gott {\it et al.}~\cite{gott01}.  The Crossing
Statistic can therefore be thought of as generalizing both the
$\chi^2$ and Median Statistics, which it approaches in different limits.

We can also look at the Crossing Statistic from another perspective: In
terms of the Gaussianity of a sample about its mean.  If an assumed
model is indeed the correct one to describe Gaussian distributed data, then the
histogram of the normalized residuals should also have a Gaussian
distribution, with zero mean and a standard deviation of
1~\cite{meanzero}. To test Gaussianity in this context one can use a variety of
different methods, including, for example, the Kolmogorov-Smirnov
test~\cite{KS-test}.  If the histogram instead exhibits significant deviation
from the a Gaussian distribution, however, then this can be used to rule out the assumed
model.  The Crossing Statistic pushes this well known idea from
statistical analysis a step further by pointing to the fact that {\it
  not only should the whole sample of residuals have a Gaussian
  distribution around the mean, but so should any continuous sub-sample}.
In our case, these sub-samples should be taken to be those residuals
within certain redshift ranges, as discussed above.
The importance of our new statement can be realized if we
look at Fig.~\ref{fig:histo} for the $T_{\rm I}$ statistic. While the overall
histogram of the normalized residuals may have a Gaussian
distribution, this does {\it not} mean that the distributions of
residuals for the data on either side of the crossing are also
Gaussian distributed.  It may be the case that the normalized
residuals to the left of the crossing point (in redshift range) contribute more to one side
of the histogram than the other, and the residuals from the other side
of the crossing point do the opposite. In essence, this is what the
$T_{\rm I}$ statistic estimates and tests. In
the case of $T_{\rm I}$, in fact, we divide the sample up into all possible
two sub-samples and we test the Gaussianity for all of them. As can be
seen in Fig.~\ref{fig:histo}, while the overall distribution seems to have a
reasonable Gaussian shape, the histogram of the normalized residuals
at $z<0.54$ has a clear shift to the right, while those at $z>0.54$
are shifted to the left. In our analysis, deviation from Gaussianity with zero
mean is calculated by derivation of $Q_1(z_1)$ and $Q_2(z_1)$ which
are, in fact, the areas under the histograms on the two sides of the
zero mean. This is a simple, but robust way, to test the
hypothesis above.

\section{Results}                        
\label{res}
\begin{figure}
\vspace{-20pt}
\includegraphics[width=95mm,height=70mm, angle=0]{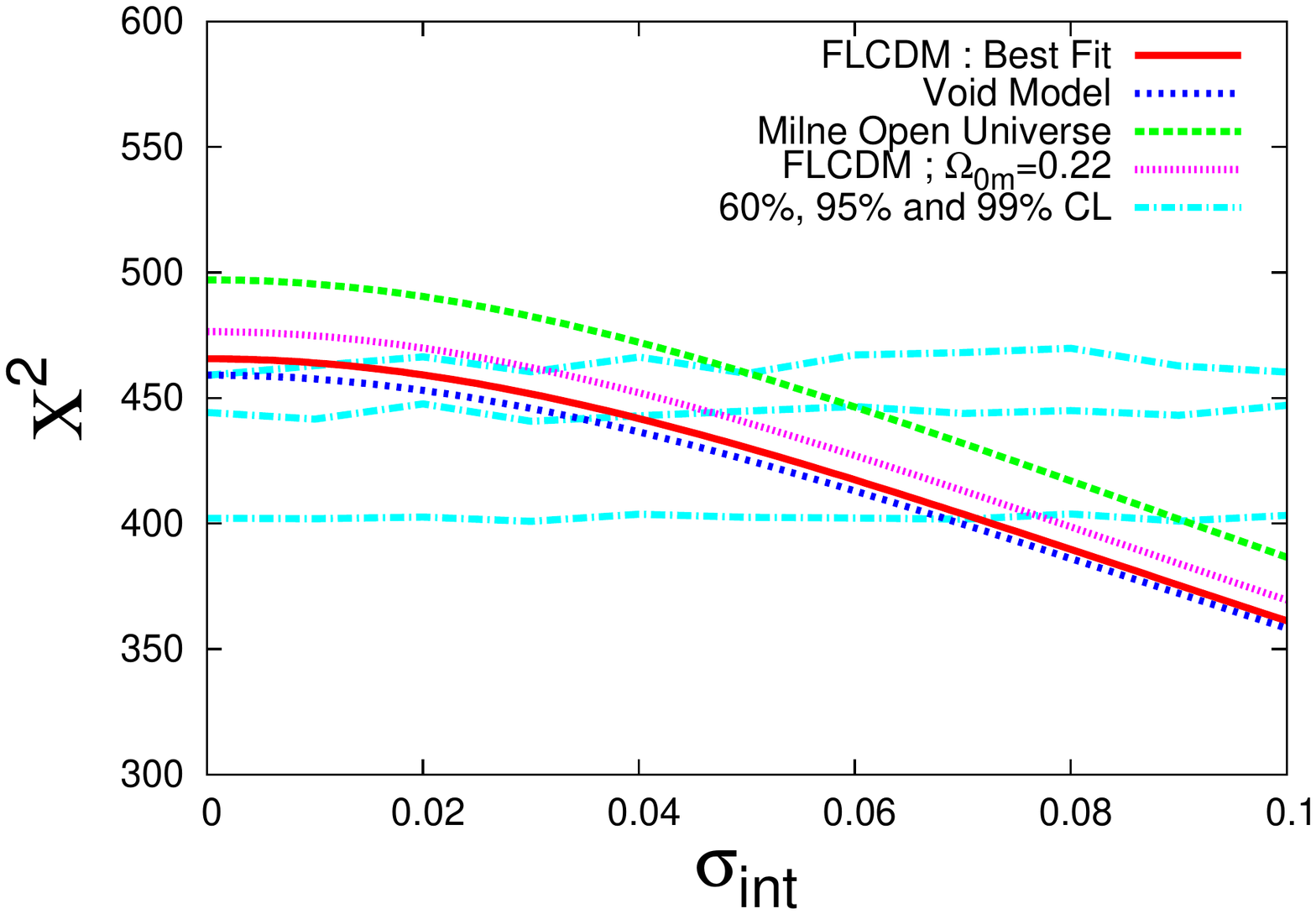}
\hfil
\hspace{-20mm}
\includegraphics[width=95mm,height=70mm, angle=0]{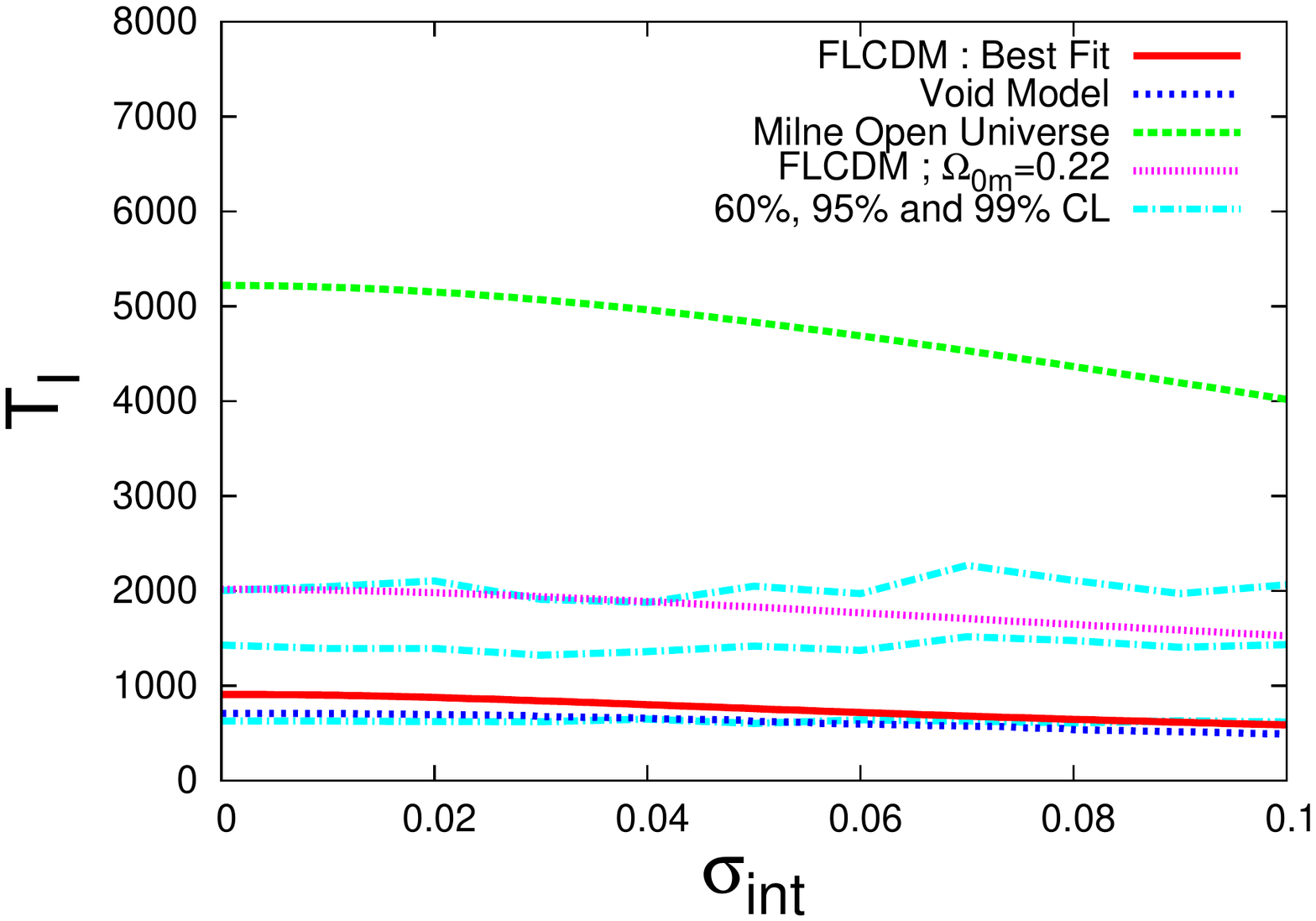}
\hfil
\includegraphics[width=95mm,height=70mm, angle=0]{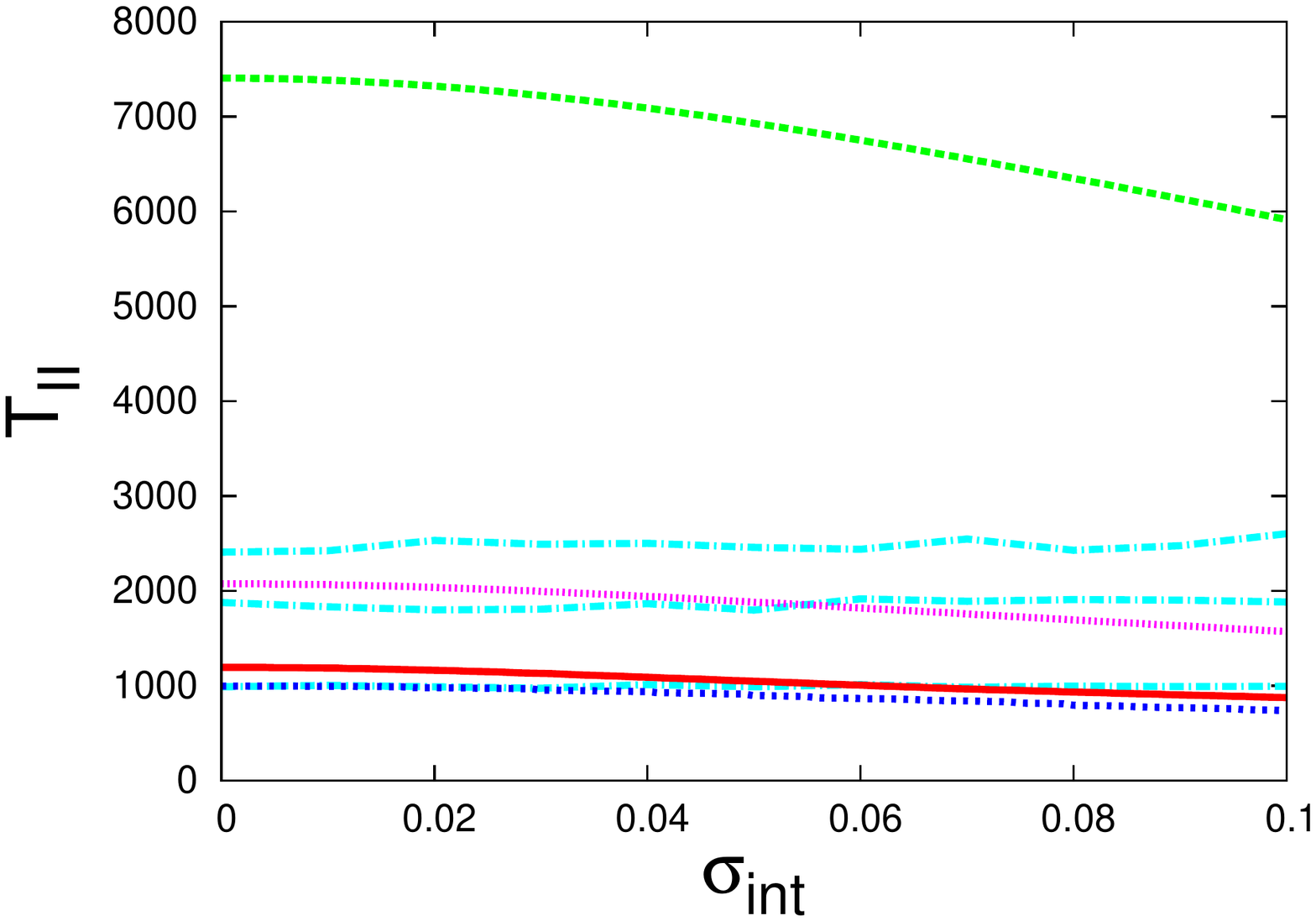}
\hfil
\hspace{-20mm}
\includegraphics[width=95mm,height=70mm, angle=0]{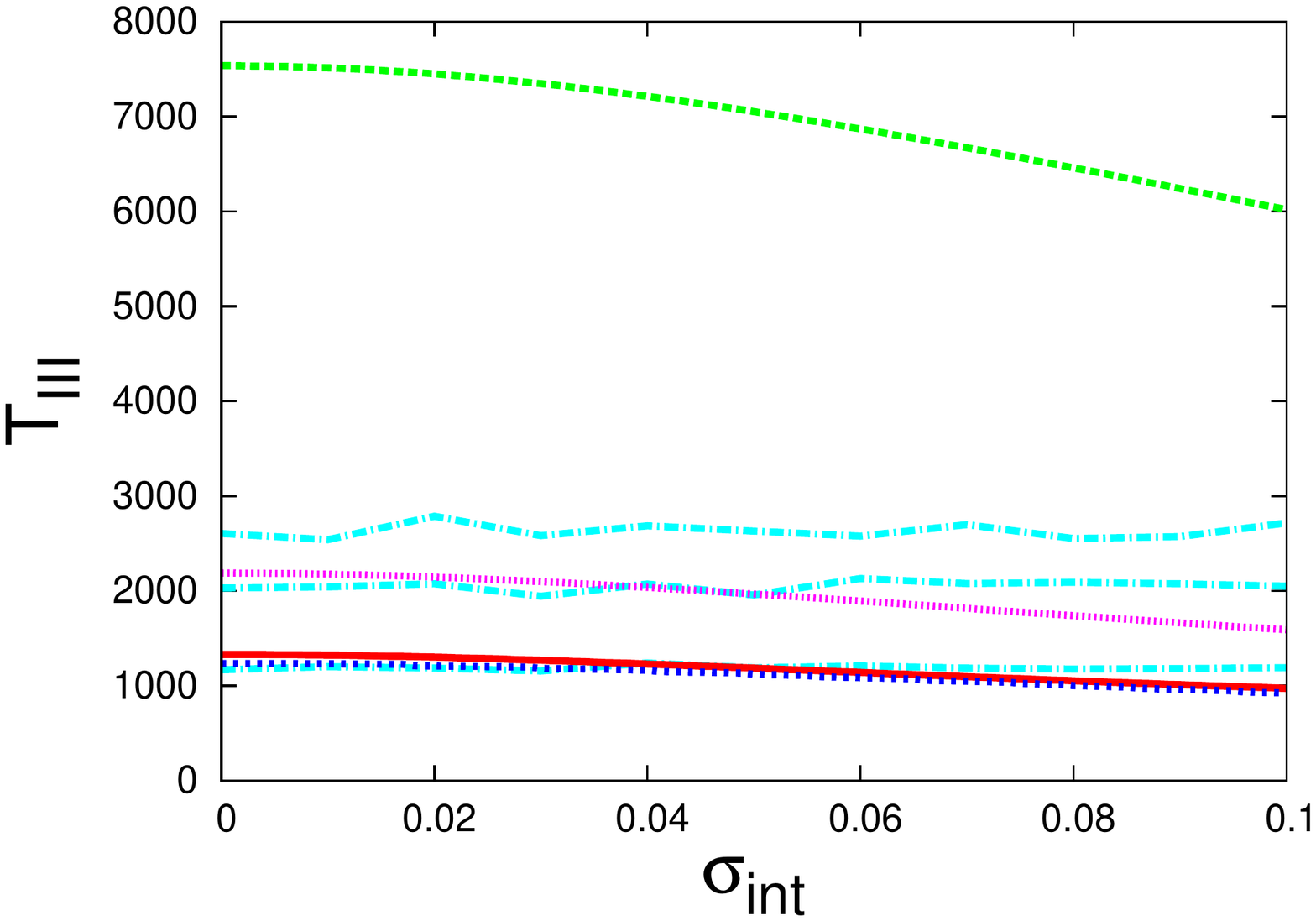}
\hfil
\caption{\small The $\chi^2$, $T_{\rm I}$, $T_{\rm II}$ and $T_{\rm III}$
  statistics for a best
  fit flat $\Lambda$CDM model (red lines), a void model (blue dashed
  lines), the Milne universe (green dashed lines) and a flat
  $\Lambda$CDM model with $\Omega_{\rm 0m}=0.22$ (pink dotted lines).
  The analyses are performed using the Constitution supernova
  data~\cite{constitution}, and by assuming various different
  additional intrinsic errors.  The confidence limits from $1000$
  Monte Carlo realizations of the error-bars are derived
  in a completely model independent manner. It can be seen
  the $\chi^2$ statistic fails to distinguish between these
  models with any degree of significance, and that by assuming
  additional intrinsic errors this statistic allows all models to be made consistent with the
  data. The $T_{\rm I}$ crossing statistic, on the other hand, rules out
  the Milne universe to more than $5\sigma$, and also the flat
  $\Lambda$CDM model with $\Omega_{\rm 0m}=0.22$ to nearly $3\sigma$, even when the amount of additional
  intrinsic error is large.} 
\label{fig:main}
\end{figure}

Now let us apply our Crossing Statistics to a suite of different models. We
will calculate $\chi^2$, $T_{\rm I}$, $T_{\rm II}$ and $T_{\rm III}$ for (i) the best fit
flat $\Lambda$CDM model (with $\Omega_{\rm 0m}=0.288$ when
$\sigma_{\rm int}=0$), (ii) a best fit asymptotically flat void
model\footnote{This
  model uses the Lema\^{i}tre-Tolman-Bondi solution of general
  relativity \cite{LTB} to model an under-density formed due to a
  Gaussian fluctuation in the spatial curvature parameter, $k$.  For
  an observer at the centre the affect of the resulting inhomogeneity
  is to create a universe that looks like it is accelerating,
  without any actual acceleration taking place.} with $\Omega_{\rm 0m}=0.28$ at the centre,
and with FWHM at $z=0.66$ when $\sigma_{\rm int}=0$, (iii) a flat $\Lambda$CDM model
with $\Omega_{\rm 0m}=0.22$, and (iv) the Milne open universe.  We use the Constitution
data set~\cite{constitution}, and vary the additional   intrinsic
error, $\sigma_{\rm int}$,  between $0$ and $0.1$ magnitudes.  In
Fig.~\ref{fig:main} we compare these
statistics with the confidence limits that result from $1000$ Monte
Carlo realizations of the error bars, for each value of $\sigma_{\rm
  int}$.  This is done in a completely model independent manner.


It can be seen from Fig.~\ref{fig:main} that the $\chi^2$ statistic (upper
panel) cannot easily be used to distinguish between the different models with a high
degree of confidence, especially if we do not know
$\sigma_{\rm int}$. Indeed, if we add $\sigma_{\rm int}=0.1$ magnitudes to the data
then all four models become a good fit, at the $60\%$ confidence
level. Alternatively, with $\sigma^2_{\rm int}=0$ all four models are outside of the $99\%$
confidence level.  This illustrates the ineffectiveness of $\chi^2$ as
a statistic for determining the goodness of fit when the errors on the
data are not well known.

The results for the one-point Crossing Statistic are shown in the second
panel from the top in Fig.~\ref{fig:main}.  In terms of this statistic
it can be seen that the best fit flat $\Lambda$CDM model and the best fit void model are now very much
consistent with the data, even with no additional intrinsic error. At
the same time, it is also clear that the Milne universe lie well outside the
$99\%$ confidence level and the flat $\Lambda$CDM model with $\Omega_{\rm 0m}=0.22$ lie well outside the
$95\%$ confidence level, even when $\sigma_{\rm int}$ is large. 
In the third and fourth panels in Fig.~\ref{fig:main} we
see the results for the two-point and three-point Crossing
Statistics, respectively. The Milne Universe remains outside the $99\%$ confidence
level in each of these, for the range of $\sigma_{\rm int}$
considered, while the flat $\Lambda$CDM model with $\Omega_{\rm 0m}=0.22$
now lies mostly within the $60$-$99\%$ confidence region.

This difference in probability of the different Crossing Statistics for the
$\Lambda$CDM model with $\Omega_{\rm 0m}=0.22$ is due to this model
having only one `crossing' with the data.  Adding extra hypothetical
crossings then has little affect on $T_i$, as the extra crossing points
all cluster around the same $z$.  A model that fits the data
better, with many crossings, however, should be expected to have $T_i$
statistics that increase with $i$.
On this basis, one can then argue that for a model to be considered consistent with the data
it must show consistency across all crossing modes. 
The point here is that if there is a significant crossing of the data
and the model, then it should show up in the Crossing Statistics as a
failure of $T_i$ to decrease sufficiently with decreasing $i$. A flat $\Lambda$CDM model
with $\Omega_{\rm 0m}=0.22$ is therefore considered non-viable at close to $3\sigma$ because of
the discrepancy in $T_{\rm I}$, even though  $T_{\rm II}$ and
$T_{\rm III}$ show some degree of consistency. 

One should notice that $\Delta \chi^2$ with respect to the best
fitting point in the parameter space can be used in deriving the
confidence only in cases where we know the
correct underlying theoretical model. If we assume an incorrect
theoretical model there will still be a best fit $\chi^2$ point in the
parameter space, and we can still define $1\sigma$, $2\sigma$ or
$n\sigma$ confidence limits, but this then has little or nothing to do with
goodness of fit or whether the assumed model is correct or not.  
While we do not know the size of the error bars, playing with the
$\sigma_{\rm int}$ can also help an incorrect model to achieve a $\chi_{\rm
  red}^2$ of one (or close to one) for the best fitting point in its
parameter space.  On the other hand, while defining the confidence limits depends on
$\Delta \chi^2$ and the degrees of freedom in the assumed parametric model, the
$\Delta \chi^2$ between two models (or even two points in the
parameter space of one model) changes with changing $\sigma_{\rm int}$
while the degrees of freedom of the assumed models are fixed.

\section{Conclusion}                        
\label{concl}

In summary, we have presented a new statistic that can be used to
distinguish between different cosmological models using their goodness
of fit with the supernova data.  Previous work on this subject has
analyzed the residuals from supernova data, and in particular has
examined pulls~\cite{kowalsky}.   In these analyses, however, the
correlations as functions of redshift have not been examined. 
Here we have included this extra information, and have shown that the different Crossing
Statistics that have been derived as a result are sensitive to the shapes and trends of the
data and the assumed theoretical model.  These statistics are in general also less
sensitive to the unknown intrinsic dispersion of the data than $\chi^2$,  
as exemplified by the fact that the consistency between a model and
a data set does not change much even when we assume large additional
intrinsic errors.  The Crossing Statistic can be used in the process of parameter
estimation, and for this purpose it can be
put in the category of {\it shrinkage estimators} \cite{shrinkage} (as
raw estimates are improved by combining them with other information in
the data set). The $\chi^2$ method, as an example of a maximum
likelihood estimator, is a very good summarizer of the data, but does not extract all
of the available statistical information.   We have shown here that by using $T_{\rm I}$,
$T_{\rm II}$ {\it etc.} we can extract more information from the data,
and use this to make more precise statements about the likelihood of
different parameters and models.

Let us now mention some of the important remaining issues that need to
be resolved in the context of the Crossing Statistic. So far, in all our
analyses, we have considered uncorrelated data. The Constitution
supernova data set~\cite{constitution} that we have used in our
analysis is, in fact, strictly uncorrelated (as all off-diagonal
elements of the correlation matrix are zero). However, in reality this
will only be approximately true, and the most recent methods of
supernova light-curve fitting results in data sets with slight correlations
between the individual data points~\cite{Union2p1}. It is an important
question as to how best to modify the Crossing Statistic to take
account of such correlations, as this would broaden the application of
the Crossing Statistic to a much wider category of problems.  Another
important issue involves comparing the Crossing Statistic with
Bayesian methods of model selection. The Crossing Statistic proposed
in this paper is by nature a frequentist approach, and is able to deal
with different models without any prior information. In contrast,
Bayesian methods require priors that play an important role in model
selection and parameter estimation.  This will complicate comparisons,
which will depend on whether we are dealing with completely unknown
phenomena (for which we have no prior information), or with phenomena where
we have some prior information available.  These issues will be
the subject of future work, and their results will reported elsewhere.

Finally, let us briefly mention the ``Three Region Test''
proposed by~\cite{Aslan_zech} that detects and maximizes the deviation between the data
and a hypothesis in three bins.  This test uses normalized residuals to test the goodness of fit
in a similar way to our Crossing Statistic, but is considerably less
general. 

The Crossing Statistic appears to us to be a promising
method of confronting cosmological models with supernovae
observations, and has the potential to
be straightforwardly generalized to other datasets where
similar problems occur.

\acknowledgments{AS thanks Subir Sarkar for his valuable suggestions, and many
useful discussions on this subject over the past few years.  We also thank
Eric Linder, Alexei Starobinsky, Istvan Szapudi and Steffen Lauritzen for their useful comments and discussions. AS acknowledges the support of the EU FP6 Marie Curie Research and
Training Network ``UniverseNet" (MRTN-CT-2006-035863) and Korea World Class University grant no. R32-10130. TC
acknowledges the support of Jesus College, Oxford.  TC and PGF both acknowledge the BIPAC.}

\appendix

\end{document}